\documentclass[journal]{IEEEtran}
\usepackage[margin=1in]{geometry}
\usepackage{threeparttable}
\usepackage{subfigure}
\usepackage{graphicx}
\usepackage{amsmath}
\usepackage{amssymb}
\usepackage{stfloats}
\usepackage{float}
\usepackage{lipsum}
\usepackage{multirow}
\usepackage{tikz}
\usepackage{hyperref}
\usepackage{comment}
\usepackage{xcolor}
\usepackage[english]{babel}
\restylefloat{figure}
\usepackage{algorithm}
\usepackage{algpseudocode}

\begin{document}

\title{Integrative Density Forecast and Uncertainty Quantification of Wind Power Generation \protect}




\author{{Jingxing Wang, Abdullah Alshelahi, Mingdi You, Eunshin Byon,~\IEEEmembership{Member,~IEEE}, and Romesh Saigal}
\thanks{Jingxing Wang, Abdullah Alshelahi, Eunshin Byon, and Romesh Saigal are with the Department of Industrial and Operations Engineering, University of Michigan,
Ann Arbor, MI 48109. Mingdi You is with Ford Motor Company, 22001 Michigan Ave, Dearborn, MI 48124. (corresponding author: shelahi@umich.edu). This work was supported by the National Science Foundation under Grants IIS-1741166, EECS-1709094 and CMMI-1536924.}}

\maketitle 

\begin{abstract}


The volatile nature of wind power generation creates challenges in achieving secure power grid operations. It is, therefore, necessary to accurately predict wind power and its uncertainty quantification. Wind power forecasting usually depends on wind speed prediction and the wind-to-power conversion process. However, most current wind power prediction models only consider portions of the uncertainty. This paper develops an integrative framework for predicting wind power density, considering uncertainties arising from both wind speed prediction and the wind-to-power conversion process. Specifically, we model wind speed using the inhomogeneous Geometric Brownian Motion and convert the wind speed prediction density into the wind power density in a closed-form. The resulting wind power density allows quantifying prediction uncertainties through prediction intervals. To forecast the power output, we minimize the expected prediction cost with (unequal) penalties on the overestimation and underestimation. We show the predictive power of the proposed approach using data from multiple operating wind farms located at different sites.

\end{abstract}



\IEEEpeerreviewmaketitle


\section{Introduction} \label{sec_intro}


Unlike traditional fossil-based energy sources, wind power generation is severely affected by stochastic weather conditions \cite{hirth2013market}, posing significant challenges in achieving secure power grid operations \cite{Bouffard2008}. Thus, accurate forecasting of wind power generation and its uncertainty quantification becomes a critical component in several decision-making processes including unit commitment, economic dispatch, and reserve determination \cite{Zhang2013}. Wind power generation forecasts have been widely investigated in the literature (e.g., \cite{Sideratos2012, Taylor2009}). Interestingly, many studies focus on generating point forecasts of wind power. However, due to the highly volatile and intermittent nature of wind power, probabilistic density forecasts become more crucial for decision-making in power system operations under large uncertainties \cite{Sideratos2012}. 

\begin{figure}[b]
  \centering
  \includegraphics[width=3.0in]{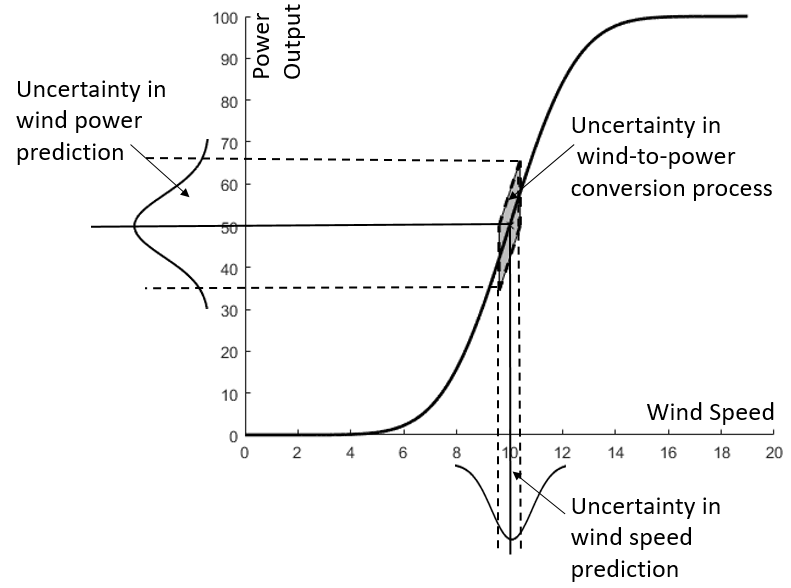}
 \caption{Uncertainties in Wind Power Output Prediction}
  \label{f1}
\end{figure}

In providing probabilistic density forecasts, prediction uncertainties should be fully recognized \cite{Choe2018}. In particular, two major uncertainty sources need to be considered. The first is the uncertainty in predicting future wind speed, whereas the second uncertainty arises when the wind speed is converted to the wind power. Such wind-to-power relationship is called \textit{the power curve}. Figure~\ref{f1} illustrates the impact of uncertainties in both wind speed forecast and conversion process on the probabilistic density prediction. Due to the nonlinearity of power curves, the predictive wind speed distribution is not linearly translated into the probabilistic characteristics of wind power prediction. Such nonlinearity causes challenges in quantifying uncertainties in wind power predictions. 

In this paper, we devise a new integrative methodology for the wind power density forecast by translating the whole predictive wind speed density into the predictive power density forecast. In particular, we formulate the wind speed as a continuous stochastic process based on the inhomogeneous Geometric Brownian Motion (GBM). The inhomogeneous GBM is flexible in capturing nonstationary and highly volatile wind characteristics. We dynamically update the time-varying parameters in the inhomogeneous GBM model with the dual Kalman Filtering in order to characterize the nonstationary nature of wind speed. By applying the Ito's lemma \cite{2009Bjork} to the stochastic power curve, we then convert the predictive wind speed density to the predictive distribution of wind  power. 

The resulting closed-form density provides a comprehensive characterization of prediction uncertainties, including predictive intervals and quantiles. 
Besides, the predictive density allows us to assign different weights on overestimating and underestimating future generation. 
For example, wind farm operators may prefer to avoid penalties due to unsatisfied demand (or unsatisfied commitment) and thus, prefer underestimation to overestimation of future wind power outputs, while others may prefer overestimation to prevent salvage of excessively generated power \cite{pourhabib2015short,HG2010_PCE}. To accommodate such unequal penalties, we formulate an optimization problem to obtain the optimal point prediction that can minimize the expected prediction cost caused by  over/underestimation, according to the operator's preference. 

We apply the proposed approach to three datasets collected from actual operating wind farms. Our implementation results indicate that the proposed approach can successfully characterize the stochastic wind power process and provide prediction results in accordance with the wind farm operator's preference.
 
The remainder of this paper is organized as follows. Section~\ref{sec-rev} reviews relevant studies. Section~\ref{sec_met} presents the proposed approach. Section~\ref{sec_re} shows the computational results on real datasets. Finally, we summarize the paper in Section~\ref{sec_con}.

\section{Literature Review} \label{sec-rev}


The fast increase in computational capabilities and data storage capacity has attracted much attention to data-driven prediction models. 
Models in this category include time-series models such as the Auto-Regressive Moving Average (ARMA), Auto-Regressive Generalized Autoregressive with Conditional Heteroscedasticity (AR-GARCH) model, and Recurrent Neural Networks (RNN). The former model is a standard approach for accounting for temporal correlation in the wind data \cite{erdem2011arma,pinson2012very}, while the latter approach allows the variance of the wind speed to vary over time by characterizing the nonstationary nature of wind conditions \cite{ZWW2014}. Another commonly used model is the persistent model. It is a simple yet effective model that assumes that the next wind speed is similar to the current speed. Despite its simplicity, the persistent model provides promising prediction accuracy at some wind sites \cite{pourhabib2015short}. 
To forecast the wind power output, the predicted wind speed is converted into wind power prediction through the power curve. Various methods are explored in the literature to estimate the power curve, including polynomial regression, splines, nonparametric regression, neural-networks and support vector machines  \cite{Yam2014,lee2015power,You2016}. Once the power curve is constructed, future wind power outputs are typically predicted by plugging the wind speed forecast into the power curve function. 

A number of prior studies have focused on providing probabilistic wind power forecasts. These studies have taken one of the following approaches. One approach is to simulate wind speed from the predictive density and convert the sampled wind speed to the power output using the power curve. For example, \cite{Taylor2009} integrates predictions generated from multiple physics-based forecast models with different scenarios via ensemble forecasts. These forecasts are then used for providing wind speed density forecast. Although this approach considers the uncertainties in predicting the wind speed, probabilistic characteristics and uncertainties in converting the wind speed to wind power are not addressed. Furthermore, as discussed in Section~\ref{sec_intro}, due to the nonlinearity in the wind-to-power conversion process, this approach does not provide the predictive wind power distribution in a closed-form. 

The second approach is to take wind speed forecast and historical wind condition as covariates (or inputs) to estimate probabilistic characteristics of wind power. Based on Neural Networks (NNs), Sideratos and Hatziargyriou \cite{Sideratos2012} estimate quantiles of future wind power, whereas prediction intervals of wind power generation are constructed in \cite{WXP2014}. In~\cite{Nielsen2006}, a linear quantile regression with spline bases is employed to estimate quantiles of the forecast errors. In many of these studies, the predictive wind speed density is not used as input, but the point wind speed forecast and/or past observations are treated as covariates. Therefore, the prediction uncertainties of wind speed are not fully captured. 


Studies in  \cite{zarate2013continuous,zarate2016construction} 
employ Brownian Motion to describe the volatility of the wind speed, where the dynamics of the wind speed are represented as stochastic differential equations (SDEs). Recently, Loukatou et al. \cite{loukatou2018stochastic} describe the continuous-time wind speed with the Ornstein-Uhlenbeck GBM model to simulate the wind power trajectory using a deterministic power curve. These studies do not take the uncertainties in the wind-to-power conversion process into consideration and are therefore limited.


In this research, we fill the gap in the literature by collectively accounting for the uncertainties arising in both wind speed prediction and stochastic power conversion process. The proposed method generates predictive density of wind power in a closed-form so that rich information can be extracted for probabilistic prediction of wind power generation. 

\section{Methodology} \label{sec_met}
In this section, we first formulate the dynamics of the wind speed process and the wind-to-power conversion process. Next, we provide an optimization framework to forecast the future wind power output based on wind farm operator's preference on over- and underestimation, and finally present the implementation procedure. 

\vspace{-.3cm}
\subsection{Modeling Wind Speed Process}\label{subsec_speed_model}
Considering the highly volatile and time-varying wind behavior, we characterize the dynamics of wind speed using the inhomogeneous GBM model \cite{oksendal2003stochastic}. 
Let $S(t)$ denote the true wind speed at time $t$. We model the stochastic process of $S(t)$ as 
\begin{equation}\label{e2.1.1}
dS(t)=\mu_S(t)S(t)dt+\sigma_S(t)S(t)dW_S(t),
\end{equation}
where $\mu_S(t)$ and $\sigma_S(t)$ capture the drift and volatility of the stochastic process, respectively, and both are time-dependent. $W_S(t)$ denotes a standard Brownian process with independent increments $\Delta W_S(t) = W_S(t + \Delta t)- W_S(t)$. These increments are normally distributed with mean 0 and variance $\Delta t$.  
 
Let $X(t)$ denote $\ln S(t)$, i.e., $X(t)=\ln S(t)$. Given the underlying dynamics of $S(t)$ in \eqref{e2.1.1}, the dynamics of $X(t)$ can be represented as 
\begin{align}\label{e2.1.2}
d[X(t)]&=\left[\mu_S(t)-\frac{1}{2}\sigma_S^2(t)\right]dt+\sigma_S(t)dW(t).
\end{align}
\noindent The detailed derivation of $X(t)$ is available in \cite[Chap. 5]{2009Bjork}. 
Solving the SDE in \eqref{e2.1.2} analytically is complicated but can be approximated by  numerical discretization. 
By applying the Wagner-Platen expansion and the Euler discretization scheme \cite{platen1999introduction} to~\eqref{e2.1.2}, we obtain 
\begin{small}
\begin{align}\label{e2.1.3}
X(t+\Delta t)=X(t)+\left[\mu_S(t)-\frac{1}{2}\sigma_S^2(t)\right]\Delta t+\sigma_S(t) \Delta W(t).
\end{align}
\end{small}
Then, it immediately follows that $X(t+\Delta t)$ in \eqref{e2.1.3} follows a normal distribution  as 
\begin{small}
\begin{equation}\label{e.1.4}
X(t+\Delta t)\sim N\left(X(t)+\left[\mu_S(t)-\frac{1}{2}\sigma_S^2(t)\right]\Delta t, \sigma_S^2(t)\Delta t\right),
\end{equation}
\end{small}
which implies that wind speed is log-normally distributed as 
\begin{align}
&\ln(S(t+\Delta t)) \label{eqn_speed_lognormal}\\
&\sim N\left(\ln(S(t))+\left[\mu_S(t)-\frac{1}{2}\sigma_S^2(t)\right]\Delta t, \sigma_S^2(t)\Delta t\right). \nonumber
\end{align}

Note that the wind speed distribution in~\eqref{eqn_speed_lognormal} characterizes the stochastic dynamics of wind speed through the time-varying parameters, $\mu_S(t)$ and $\sigma_S(t)$. To estimate $\mu_S(t)$ and $\sigma_S(t)$, one should use wind measurements collected from a meteorological tower or turbine anemometers. However, the collected wind speed may have measurement errors and/or can be perturbed by disturbances such as wake effects \cite{You2016}. Therefore, the true wind speed $S(t)$ is unobserved in practice.   
To incorporate such errors and disturbances, assume that the measured wind speed is a linear function of the unobserved true   speed. Let $WS(t)$ denote the measured wind speed at time $t$ and $Y(t)=ln (WS(t))$. Letting $X(t)(=ln (S(t))$ be a state variable perturbed by a normally distributed error term $z \sim N(0,\sigma^2_z)$ as follows.
\begin{align} \label{eqn-state}
Y(t) =X(t)+z.
\end{align}

Note that the dynamics of $X(t)$, governed by the linear SDE representation in~\eqref{e2.1.3}, can be rewritten as 
\begin{equation}\label{e2.1.4}
X(t+\Delta t) =X(t) + A \:\theta(t) +w(t), \\
\end{equation}
where $A=\left(\Delta t, - \frac{1}{2} \Delta t\right)$, $\theta(t)=\big(\mu_S(t),\sigma_S^2(t)\big)^T$, and $w(t) \sim N(0, \sigma_S^2(t)\Delta t)$ is the process noise.

The equations in~\eqref{eqn-state} and~\eqref{e2.1.4} together represent the linear state space model. Among several ways to estimate the model parameters in the linear state space model, we employ the Kalman filter due to its flexibility and strong performance in many applications \cite{haykin2004kalman,sun2014building}. 
In particular, we employee the dual Kalman filtering procedure to estimate parameter vector $\theta(t)$ and state $X(t)$ \cite{wan1997dual}.   
To model the time-varying parameter $\theta(t)$, we assume that it drifts according to a two-dimensional Gaussian random walk process with covariance $Q$, i.e., 
\begin{equation}\label{e2.1.8}
\theta(t+\Delta t)=\theta(t) + \epsilon,
\end{equation}
where $\epsilon\sim  N(0,Q)$. We include the detailed procedure to update the parameters $\theta(t) $ and state $X(t)$ in Appendix. 



\vspace{-.3cm}
\subsection{Modeling Wind-to-Power Conversion Process}\label{subsec_power_model}
This section discusses how to convert the wind speed dynamics into the dynamics of wind power process. The relationship between the wind speed and the power generation can be quantified by the power curve function. Let $F(t,S(t))$ denote the power curve at time $t$, given the speed $S(t)$. Here, $F(t,S(t))$ can represent the power curve from a whole wind farm or a stand-alone turbine. 

We model the power curve function $F(t,S(t))$ as a function of $t$ (as well as $S(t)$) to incorporate the time-varying nature of power generation efficiency. This is because, in addition to the wind speed, the wind power output depends on many other environmental factors such as wind direction, humidity, and ambient temperature \cite{lee2015power}. Moreover, turbines' age and degradation states of their components (e.g., blade, gearbox) also affect the generation efficiency. Including all of these additional factors, if not impossible, would make the power curve model overly complicated, and more importantly, it also needs to characterize the dynamics of each factor, as we did for wind speed in Section~\ref{subsec_speed_model}. Instead, we consider the power curve as a function of wind speed only and let the power curve function itself time-varying. Our approach in modeling the power curve is flexible enough to  employ a time-invariant power curve that only depends on wind speed; in this case, the power curve function can be simply reduced to $F(t,S(t))=F(S(t))$.

In  modeling $F(t,S(t))$, any type of functions, e.g., parametric, semi-parametric such as splines \cite{lee2013}, or nonparametric function  \cite{lee2014kernel,byon2015adaptive}, can be employed as long as $F(t,S(t))$ satisfies some weak conditions. Suppose that the power curve function $F(t,S(t))$ is differentiable over $t$ and $S(t)$ and twice differentiable over $S(t)$. The power output $P(t)$ at time $t$ is given by 
\begin{align}P(t)=F(t,S(t))+e(t), \label{eqn-power}
\end{align}
where $e(t)$ is a random noise in the wind-to-power conversion process. 
We assume that $\Delta e(t)=e(t+\Delta t)-e(t)$ follows the normal distribution with mean $0$ and variance $\sigma_F^2 F_S(t,S(t))\Delta t$, 
where $F_S(t,S(t))$ represents the first derivative of $F(t,S(t))$ over $S(t)$. 
In formulating the noise variance, we include $F_S(t,S(t))$, because the power conversion variability tends to be high when the power curve changes rapidly, which is mostly in the mid-speed range. For notational brevity, we will use $F_S$ as an abbreviation of $F_S(t,S(t))$ in the subsequent discussion. 


In the subsequent sections, we first model the dynamics of the wind power process with any power curve function  $F(t,S(t))$. Then, we derive the dynamics with a specific form for $F(t,S(t))$ to illustrate our approach.

\subsubsection{Dynamics of Wind Power Process with General Power Curve Function}

Given the wind speed process $S(t)$   in~\eqref{e2.1.1}, the wind power process also follows the inhomogeneous GBM and its dynamics is modeled by 
%
\begin{align}
dP(t)
&=\mu_P(t)P(t)dt+\sigma_P(t) P(t) dW_P(t)\label{e123}
\end{align}
with
\begin{align}
\mu_P(t)&=\frac{F_t+\mu_S S F_S+\frac{1}{2}\sigma_S^2S^2F_{SS}}{P(t)}, \label{eqn_mu_p}\\
\sigma_P(t)&=\frac{\sqrt{\sigma_S^2 S^2 F_S^2+\sigma^2_F F_S}}{P(t)},   \label{eqn_sigma_p}
\end{align}
where $W_P(t)$ denotes a standard Brownian process, $F_t$ represents the first derivative of $F$ over $t$, and $F_{SS}$ is the second derivative of $F$ over $S$. Also, $S$, $\mu_S$, and $\sigma_S$ denote $S(t)$, $\mu_S(t)$, and $\sigma_S(t)$ in~\eqref{e2.1.1}, respectively. We derive ~\eqref{e123}-\eqref{eqn_sigma_p} using Ito's Lemma in \cite[Chap. 4]{2009Bjork}.

Note that $\mu_P(t)$ and $\sigma_P(t)$ in~\eqref{eqn_mu_p} and~\eqref{eqn_sigma_p}, respectively, depend on the parameters in $S(t)$ (i.e., $\mu_S$, $\sigma_S$) and the  power curve related functions (i.e., $F_t, F_S, F_{SS}$). This result indicates that the stochastic dynamics of wind speed $S(t)$, together with  the power curve function, is translated into the dynamics of power generation $P(t)$. 

Following the similar procedure in~\eqref{e2.1.1}-\eqref{eqn_speed_lognormal}, one can derive the distribution of wind power in a closed-form. Specifically, the power output $P(t+\Delta t)$ at time  $t+\Delta t$ is log-normally distributed as
\begin{align} 
& \ln(P(t+\Delta t))\label{powerlognormal}\\
&\sim N\left(\ln(P(t))+\left[\mu_P(t)-\frac{1}{2}\sigma_P^2(t)\right]\Delta t, \sigma_P^2(t)\Delta t \right). \nonumber  
\end{align}


\subsubsection{Dynamics of Wind Power Process with Nonparametric Power Curve Function} \label{sec-pc-nonpara}

As discussed earlier, the power curve  $F(t,S(t))$ can be flexibly modeled using various functional forms. To illustrate, we employ the nonparametric adaptive power curve model \cite{byon2015adaptive} in our analysis. We explain only an outline of the nonparametric adaptive model in this study. For more detailed procedure, the reader is referred to  \cite{byon2015adaptive}. 

In the nonparametric approach, the input $S(t)$ is mapped into a feature space through a nonlinear mapping $S(t) \rightarrow \phi(S(t))$. Then $P(t)$ can be modeled by 
\begin{equation}
P(t)=F(t,S(t))+e(t)=\omega_t^T\phi(S(t))+e(t),
\end{equation}
where $\omega_t$ is a nonparametric regression coefficient vector at period $t$. 

The coefficient vector $\omega_t$ is time-varying, so that the power curve $F(t,S(t))$ can be updated whenever a new sample is observed. Suppose that  $\omega_{t-\Delta t}$ was estimated by $\hat{\omega}_{t-\Delta t}$ at time $t-\Delta t$ and we obtain newly observed data at time $t$. Then we estimate $\omega_t$ by solving the following optimization problem.
\begin{align}
\min L&=\frac{1}{2}\| \omega_t-\hat{\omega}_{t-\Delta t}\|^2+\frac{1}{2}\gamma e(t)^2\label{eqn-pc-obj} \\
s.t.& \quad P(t)=\omega_t^T\phi(S(t))+e(t). \label{eqn-pc-con}
\end{align}
Here the first term in the objective function represents the change of the coefficient from $t-\Delta t$ to $t$. The second term regularizes the amount of update with the regularization parameter $\gamma$, balancing the coefficient change and  quality of model fitting. 

Let $k(S(t_i),S(t_j))$ denote the inner product of $\phi(S(t_i))$ and $\phi(S(t_j))$, i.e., $k(S(t_i),S(t_j))=\phi(S(t_i)),\phi(S(t_j)))$, which is called  a kernel function. Suppose there are $n$ observations up to time $t$. Then $F(t,S(t))$ is updated by 
\begin{equation}
\hat{F}(t,S(t))=\sum_{i=1}^{n}\lambda_i k(S(t),S(t-(n-i)\Delta 
t)), \label{eqn-pc-est}
\end{equation} 
 where $\lambda_i$ is Lagrange multiplier corresponding to the equality constraint in~\eqref{eqn-pc-con}. Among many choices of the kernel function, we employ the  Gaussian kernel due to its flexibility,  

Then the estimated power curve, $\hat{F}(t,S(t))$ in~\eqref{eqn-pc-est}, can be plugged into the predictive distribution for $P(t+\Delta t)$ in~\eqref{powerlognormal}. Specifically, to estimate $\mu_P(t)$ and $\sigma_P(t)$ in~\eqref{eqn_mu_p} and~\eqref{eqn_sigma_p}, respectively, we need to estimate $F_t$, $F_S$, $F_{SS}$ and $\sigma_F$.    First, $F_t$ can be  estimated by taking the finite difference as 
\begin{align}
\hat{F}_t=\frac{\partial F}{\partial t}  \notag &=\frac{\hat{F}(t,S(t))-\hat{F}(t-\Delta t,S(t))}{\Delta t} \\ &=\frac{\lambda_tk(S(t),S(t))}{\Delta t}.
\end{align}
Next,  $F_S$ and $F_{SS}$, which  are partial derivatives of $F$ over $S(t)$, respectively, can be  estimated by
\small
\begin{align}
\hat{F}_S &=\frac{\partial F}{\partial S}=\sum_{i=1}^n\lambda_i\dfrac{\partial k(S(t),S(t-(n-i)\Delta 
t))}{\partial S(t)}\notag\\
&=\sum_{i=1}^n\lambda_ik(S(t),S(i\Delta t))\left(-\frac{S(t)-S(t-(n-i)\Delta 
t)}{\delta}\right),
\end{align}
\normalsize
 and 
\begin{align}
\hat{F}_{SS}&=\frac{\partial^2 F}{\partial S^2}=\sum_{i=1}^n\lambda_i\dfrac{\partial^2 k(S(t),S(t-(n-i)\Delta 
t))}{\partial S^2(t)}\notag\\
&=\sum_{i=1}^n \lambda_ik(S(t),S(t-(n-i)\Delta 
t)) \notag\\ &\cdot   \left(\frac{(S(t)-S(t-(n-i)\Delta t))^2}{\delta^2}-\frac{1}{\delta}\right).
\end{align}


 
Finally,  to estimate $\sigma_F$ in $\Delta e_t\sim N(0,\sigma_F^2 F_S(t,S(t))\Delta t)$, we use the sample standard deviation  with the first $N_0$ data points as follows.
\small
\begin{equation}\label{sigma_F}
\hat{\sigma}_F=\sqrt{\frac{1}{N_0-2}\sum_{i=2}^{N_0}\left(\frac{\Delta P(i\Delta t)-\Delta \hat{F}(i\Delta t,S(i\Delta t))}{\sqrt{\hat{F}_S(i\Delta t,S(i\Delta t))\Delta t}}\right)^2},
\end{equation}
\normalsize
where
\begin{align}
\Delta P(i\Delta t)&=P(i\Delta t)-P((i-1)\Delta t)\\
\Delta \hat{F}(i\Delta t,S(i\Delta t))
&=\hat{F}(i\Delta t,S(i\Delta t)) \notag \\ &-\hat{F}((i-1)\Delta t,S((i-1)\Delta t))
\end{align}

By plugging the estimated parameters, $\hat{F}$,  $\hat{F}_t$ $\hat{F}_S$, $\hat{F}_{SS}$ and $\hat{\sigma}_F$ into \eqref{eqn-pc-est}-\eqref{sigma_F}  to $\mu_P(t)$ and $\sigma_P(t)$ in~\eqref{eqn_mu_p} and~\eqref{eqn_sigma_p}, we obtain the predictive distribution of power at $t+\Delta t$ in~\eqref{powerlognormal}. Recall that other parameters associated with wind speed dynamics, i.e., $\mu_S$ and $\sigma_S$, are estimated from the dual Kalman filtering process discussed in Section~\ref{subsec_speed_model}.

\vspace{-.3cm}
\subsection{Uncertainty Quantification and Wind Power Prediction}\label{subsec_pred}

The closed-form predictive distribution of wind power output in~\eqref{powerlognormal} provides comprehensive information to characterize prediction uncertainties such as the prediction interval and quantiles. Following the procedure discussed in \cite{Dahiya1982}, the $(1-\beta)100\%$ prediction interval for the power generation at time $t+\Delta t$ is given by  
\begin{equation}\label{eqn-pred-interval}
\left[exp(\mu'+\sigma' A),exp(\mu'+\sigma' B)\right]
\end{equation}
where $\mu'=\ln(P(t)+\left[\mu_P(t)-\frac{1}{2}\sigma_P^2(t)\right]\Delta t$ and $\sigma'=\sigma_P(t)\sqrt{\Delta t}$, and $A$ and $B$ are the solution of 
\begin{equation}
\begin{cases}
\Phi(B)-\Phi(A)=1-\beta,\\
A+B=-2\sigma'.
\end{cases}
\end{equation}
Here $\Phi(\cdot)$ denotes the cumulative distribution function of a standard normal distribution.
 
 The $\alpha$-quantile $Q_{\alpha}$ such that $Pr(P(t+\Delta t)\le Q_\alpha)=\alpha$ is obtained by 
 \begin{equation}
 Q_\alpha=exp(\mu'+\sigma'\Phi^{-1}(\alpha)).\label{eqn_quantile}
 \end{equation}
In particular, the median of $P(t+\Delta t)$ is given by $exp(\mu')$ for $\alpha=0.5$.

The quantile information is critical in determining the prediction value. In time series analysis, quantities that represent a central tendency, e.g., mean or median, are typically used as a point forecast. Such forecast might not be accurate when the cost of underestimation and overestimation are different, as in wind power operations~\cite{pourhabib2015short, AlShelahi2019}. Given the quantile, the power is flexibly estimated by penalizing under/overestimation differently.

Let $p$ denote the predicted power output at time $t+\Delta t$. Let $f(x)$ is the probability density function (pdf) of the log-normal distribution described in \eqref{powerlognormal} of the power output at $t+\Delta t$. The expected amount of underestimation and overestimation, denoted by $u(p;t+\Delta t)$ and $o(p;t+\Delta t)$, respectively, are given by 
\begin{align}
u(p;t+\Delta t)&=E_{P(t+\Delta t)}[\max\{0,P(t+\Delta t)-p\}]\notag\\
&=\int_p^{+\infty}xf(x)dx,\label{e.under}\\
o(p;t+\Delta t)&=E_{P(t+\Delta t)}[\max\{0,p-P(t+\Delta t)\}]\notag\\
&=\int_{-\infty}^p xf(x)dx.
\end{align}

To predict the power output, one can minimize the expected cost due to possible under/overestimation. Therefore, the optimal $p$, denoted by $p^*$, is obtained by solving the following unconstrained optimization problem.
\begin{equation}\label{optimization}
p^*=\operatorname*{arg\,min}_p \left(\alpha \cdot u(p;t+\Delta t)+(1-\alpha) \cdot o(p;t+\Delta t) \right)
\end{equation}
where $\alpha\in[0,1]$ represents the penalty to the underestimation. When the underestimation (overestimation) is more costly, $\alpha$ greater (less than) than 0.5 can be used. It is straightforward to show that the optimal solution of \eqref{optimization}  is the $\alpha$-quantile in~\eqref{eqn_quantile} \cite{pourhabib2015short}. In other word, the solution of~\eqref{optimization} is given by $p^*=Q_\alpha$. 

\subsection{Implementation Details}\label{subsec_imp}

 In our implementation, we divide each wind farm dataset into training and testing sets. The training set includes $N_0$ observations in the first 70\% samples of the whole dataset obtained from each wind farm. The parameters $\sigma_S(t)$ and $\mu_S(t)$ in the wind speed process, the error parameters ($\sigma^2_z$ in~\eqref{eqn-state} and $Q$ in~\eqref{e2.1.8}) in the dual Kalman filtering, and the power curve are initialized using the $N_0$ observations in the training set. In particular, to set the error parameters in the kalman filtering, we apply the validation technique to the $N_0$ data points and choose the values that minimize the prediction error \cite{friedman2001elements}.  Moreover, considering that $ln(S(t+\Delta t))$ is normally distributed as shown in~\eqref{eqn-power}, we use the sample mean and sample standard deviation of the measured wind speeds to initialize $\mu_S(N_0)$ and $\sigma_S(N_0)$ (see the lines \#6-\#8 in the algorithm).
 
The testing set contains the remaining 30\% samples and is used for evaluating the prediction performance in each wind farm. In this prediction step we update (or filter)  the model parameters whenever a new observation is obtained. In Algorithm~1, $\mu_S(t+1\mid t)$, $\sigma_S(t+1\mid t)$, and $S(t+1\mid t)$  in line  \#12 denote the prior estimates of $\mu_S(t+1)$, $\sigma_S(t+1)$, and $S(t+1)$, respectively, from the Kalman filtering, whereas $\mu_S(t+1\mid t+1)$, $\sigma_S(t+1\mid t+1)$ and $S(t+1\mid t+1)$ in the filtering step (lines  \#15-\#18), correspond to their posterior estimates after observing wind speed $WS(t+1)$ and power $P(t+1)$ at time $t+1$; more detailed dual Kalman filtering procedures are included in Appendix. 

\section{Case Studies} \label{sec_re}

We apply the proposed approach to real datasets collected from three operating wind farms,  WF1, WF2, and WF3, summarized in Table~\ref{tabel_wf}. Due to the data confidentiality required by the data providers, detailed information regarding each wind farm is omitted.  Each dataset includes wind measurements and power outputs from the whole wind farm. In all wind farms, the power outputs are scaled to $[0,100]$. 

\begin{table}[ht] 
\centering
\caption{Wind Farms Information}\label{tabel_wf}
\begin{tabular}{   c  || c || c || c}\hline \hline
 {\textbf{Dataset}} & WF1 & WF2 & WF3 \\\hline \hline
 { {Terrain}} &  offshore &  land-based &  onshore \\\hline
  { {Number of turbines}} &  about 35 &  240+ & about 10\\\hline
 {Total data size} &  1000 & 1000 & 650\\\hline
  {Temporal resolution} &  10 minute& 10 minute & 10minute\\\hline \hline
 \end{tabular}
\end{table}

\vspace{-.5cm}
\subsection{Implementation Results}

\begin{figure}[ht]
  \centering
  \includegraphics[width=3.2in]{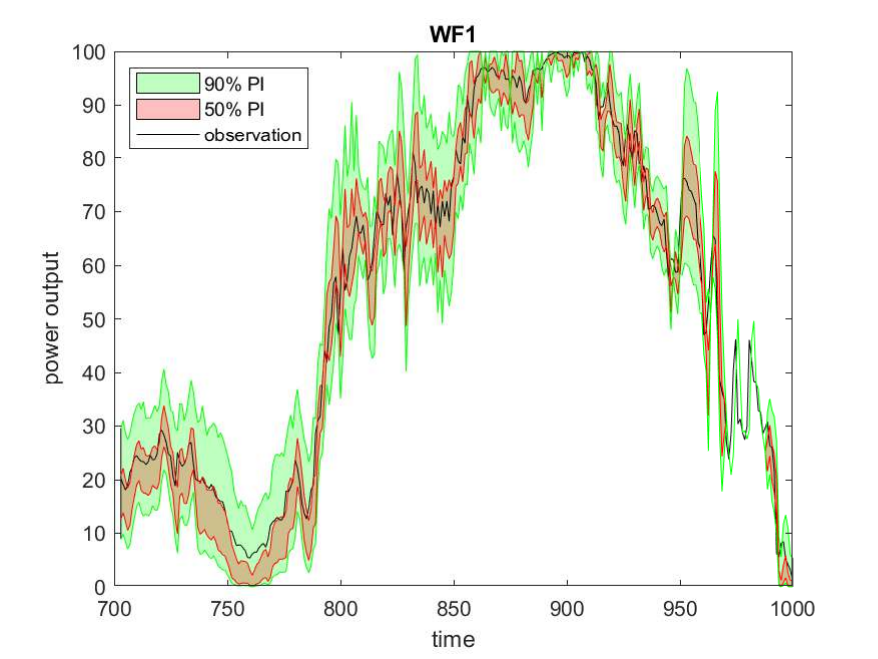}
  \caption{Power Output Prediction Intervals on WF1 Dataset}
  \label{PI} 	
 
\end{figure}

    
Figure~\ref{PI} depicts the $50\%$ and $90\%$ prediction intervals in WF 1 testing set. Note that the upper bound is capped at 100 (the maximum normalized power output). The majority of the observations fall inside the prediction intervals, indicating that our approach can successfully capture the uncertainties. We can also observe that in general the more volatile the power output (i.e., when the power output changes rapidly), the wider the prediction intervals. For example, when $t$ is about 950, the power output changes rapidly and the prediction intervals are wider, which represents larger prediction uncertainties. On the other hand, when the output is less volatile, e.g., when $t$ is between $860$ and $870$, we obtain narrower intervals. We observe similar patterns in other wind farms but omit the results to save space. 

\vspace{-.5cm}
\subsection{Comparison with Alternative Methods}

We compare our approach with other alternative methods, including the persistent model,  ARMA, AR-GARCH model, NN with long short-term memory (LSTM) layers. 
In both ARMA and AR-GARCH methods, the wind speed   is assumed to follow a normal distribution. In implementing ARMA, AR-GARCH, and LSTM NN, we use built-in functions in Matlab. 

Each model order (number of parameters) is chosen such that the Bayesian information criterion (BIC) is minimized  To determine the structure of LSTM, including the number of layers and the number of neurons, we apply the following validation technique. The networks are trained using 50\% of the whole data set in each wind farm and the prediction performance is evaluated using the validation set consisting of 20\% of the data set. We choose the best network structure with the lowest prediction error in the validation set. 
The models in these alternative methods are re-trained when a new observation is obtained. Once the wind speed at time $t+\Delta t$ is predicted using these approaches, the predicted wind speed is plugged into the power curve to get $\hat{P}(t+\Delta t)$. In all four methods, we apply the same adaptive non-parametric power curve discussed in Section~\ref{subsec_power_model}.

We evaluate the prediction performance with different penalties on the overestimation and underestimation. In the proposed approach, we use the $\alpha$-quantile of the predictive power output density as discussed in Section~\ref{sec_met}. For fair comparison, in ARMA and AR-GARCH, we also use the $\alpha$-quantile of their predictive wind speed densities and plug the resulting $\alpha$-quantile estimates to the power curve \cite{pourhabib2015short}. Note that the forecast values do not change with different $\alpha$ values in the persistent and LSTM NN method, because they do not provide predictive densities but  only provide point predictions.  

We measure the prediction quality with unequal penalties using power curve error (PCE)~\cite{HG2010_PCE} defined as
\begin{align}\label{eqn_pce}
&PCE(P(t),\hat{P}(t)) =  \\
&\begin{cases}
      \alpha(P(t)-\hat{P}(t)), & \text{if }\hat{P}(t) < P(t) \\
       (1-\alpha)(\hat{P}(t)-P(t)), & \text{otherwise.}
    \end{cases} \nonumber
\end{align} 
where  $P(t)$ is the observed power at time $t$ and $\hat{P}(t)$ is its predicted power from each method.  

\begin{figure}[!ht]
        \centering
            \includegraphics[width=2.3in]{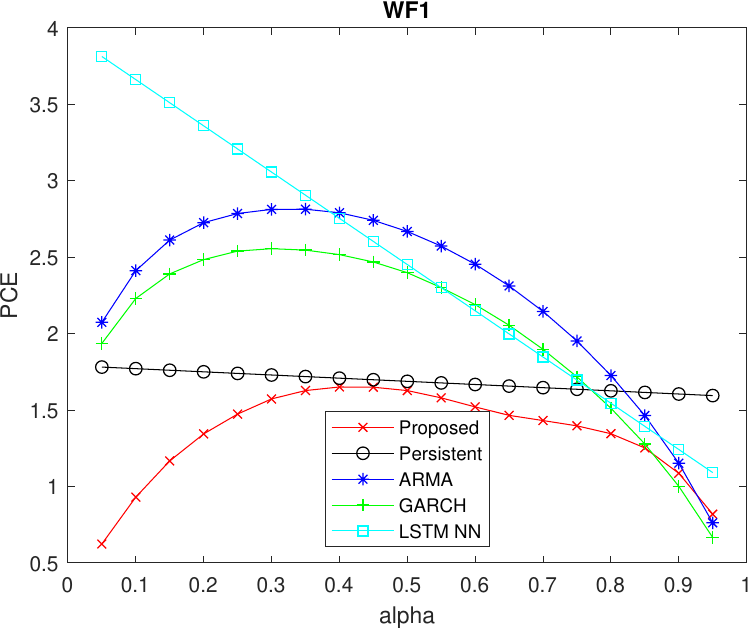}
            \includegraphics[width=2.3in]{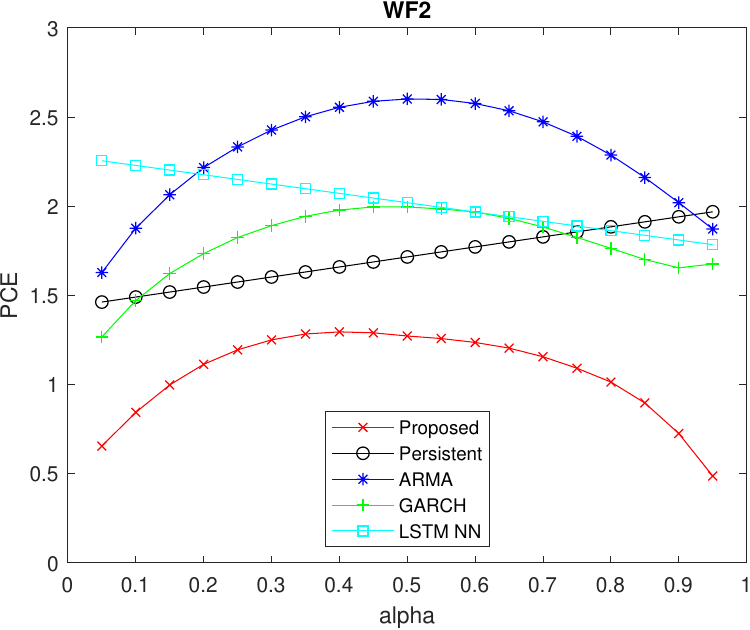}
            \includegraphics[width=2.3in]{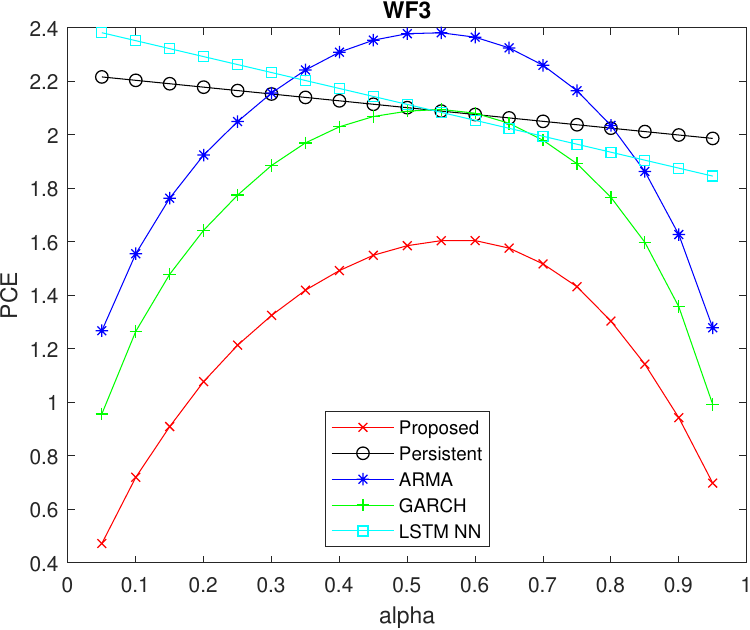}
        \caption
        {Average power curve errors in the testing set} 
        \label{PCE}
    \end{figure}

Table~\ref{tabel_PCE} summarizes the average PCE from each method for three $\alpha$ values in the testing set.  
Figure~\ref{PCE} further shows the average PCE over $\alpha\in[0,1]$. The AR-GARCH generates lower PCEs than ARMA, because it takes time-varying variance of wind speed into consideration. But PCEs from AR-GARCH are still higher than the proposed approach in all datasets. The LSTM NN also generates higher PCEs than the proposed approach. Our approach consistently produces the lowest PCEs in all cases, indicating that our approach is superior in reflecting wind farm operators' prediction preference on overestimation and underestimation.

\begin{table}[!ht]
\centering
\caption{Average Power Curve Error in the testing set. Boldfaced values indicate the best performance.}\label{tabel_PCE}
\begin{tabular}{ c | c  ||c || c || c}\hline \hline
$\alpha$ & Approach & WF1 & WF2 &WF3\\\hline \hline
\multirow{6}{*}{$ {0.27}$}& Proposed  Approach &  \textbf{1.52} & \textbf{1.22} & \textbf{1.26} \\\cline{2-5}
& Persistent & 1.74 & 1.59 & 2.16 \\\cline{2-5}
& ARMA & 2.80 & 2.37 & 2.09 \\\cline{2-5}
& AR-GARCH & 2.55 & 1.85 & 1.82\\\cline{2-5}
& LSTM NN & 3.15 & 2.14 & 2.25\\
\hline

\multirow{6}{*}{$ {0.5}$}& Proposed  Approach& \textbf{1.63} & \textbf{1.27} & \textbf{1.59} \\\cline{2-5}
& Persistent & 1.69 & 1.72 & 2.10 \\\cline{2-5}
& ARMA & 2.67 & 2.60 & 2.38 \\\cline{2-5}
& AR-GARCH & 2.40 & 2.00 & 2.09\\\cline{2-5}
& LSTM NN & 2.45 & 2.02 & 2.11\\
\hline

\multirow{6}{*}{$ {0.73}$}& Proposed  Approach& \textbf{1.41} & \textbf{1.12} & \textbf{1.47} \\\cline{2-5}
& Persistent & 1.64 & 1.85 & 2.04 \\\cline{2-5}
& ARMA & 2.03 & 2.43 & 2.21 \\\cline{2-5}
& AR-GARCH & 1.79 & 1.85 & 1.93\\\cline{2-5}
& LSTM NN & 1.76 & 1.90 & 1.98\\
\hline\hline
\end{tabular}
\end{table}

\vspace{-.7cm}
\section{Summary} \label{sec_con}
We present a new integrative methodology for predicting the wind power density. 
The proposed approach accounts for uncertainties in wind speed process and wind-to-power conversion process and provides rich information for the probabilistic forecast through its closed-form prediction density. The closed-form density allows us to extract diverse information and to determine forecast, depending on the wind farm operator's preference on the overestimation and underestimation of future wind power outputs. This framework can minimize the overall costs associated with prediction errors.

We believe that our approach could potentially benefit power grid operations. In the future, we plan to incorporate our prediction results into the optimization framework for solving decision-making problems such as economic dispatch. We also plan to apply the approach to predict the mechanical and structural load responses in the wind turbine system for the reliability analysis and maintenance optimization \cite{KB2017, Choe2015}. The proposed methodology is also applicable to other engineering systems subject to nonstationary operating conditions, such as solar power systems \cite{CG2016}. \\

 \vspace{-.5cm}
\bibliographystyle{WileyNJD-AMA}

\bibliography{WindFarmOptions}

\section{\textbf{Appendix}}

\noindent \emph{\textbf{Dual Kalman Filtering Procedure}}\label{app2} 

Recall that the parameter vector is $\theta(t)=[\mu_S(t), \sigma^2_S(t)]^T$ and state is $X(t)$.  We use $\theta_2(t)$ for $\sigma^2_S(t)$. Let $\hat{X}(t \mid t)$ and $\hat{X}(t + \Delta t \mid t)$ denote the posterior and prior estimates of state variable $X(t)$ with their associated estimation error variances $P_X(t \mid t)$ and $P_X(t + \Delta t \mid t)$, respectively. Similarly, $\hat{\theta}(t \mid t)$ and $\hat{\theta}(t + \Delta t \mid t)$, respectively, denote the posterior and prior estimates of the parameter vector $\theta(t)$ and $P_{\theta}(t \mid t)$ and $P_{\theta}(t + \Delta t \mid t)$ represent the corresponding estimation error covariance matrices. We let $K_X(t)$ and $K_{\theta}(t)$ denote the Kalman gain associated with state and parameters filters at time $t$, respectively. Then the dual Kalman filtering proceeds as follows:
\begin{itemize}
\item Parameters prediction:
\footnotesize\begin{align} 
\hat{\theta}(t +\Delta t\mid t)&= \hat{\theta}(t \mid t),\nonumber \\
P_\theta (t+\Delta t \mid t)&= P_\theta (t\mid t) + Q. \nonumber
\end{align}
\item State prediction:
\footnotesize\begin{align} 
\hat{X}(t+\Delta t \mid t )&= \hat{X}(t \mid t) + A \: \hat{\theta}(t+\Delta t \mid t), \nonumber\\
P_X (t+\Delta t \mid t)&= P_X (t \mid t) + \Delta t \: \hat{\theta}_2(t+\Delta t \mid t).\nonumber
\end{align}
\item State filtering:
\footnotesize\begin{align}
K_X(t+\Delta t)&=P_X (t+ \Delta t \mid t) \: \:\big[P_X (t+\Delta t \mid t)+ \sigma^2_z]^{-1},\nonumber\\ \nonumber
\hat{X}(t+\Delta t \mid & t+\Delta t) = \hat{X}(t+\Delta t \mid t) \notag\\
&+ K_X(t+\Delta t) \: \big[Y(t+\Delta t) - \hat{X}(t+\Delta t \mid t) \big],\nonumber\\
P_X (t+\Delta t \mid & t+\Delta t) = \big[I-K_X(t+\Delta t)\big] \: P_X (t+\Delta t \mid t). \nonumber
\end{align}
\item Parameters filtering:
\footnotesize\begin{align}
K_\theta(t+\Delta t)&= \nonumber\\
&\hspace{-1cm}P_\theta (t+\Delta t \mid t) \: A^T \: \big[A \: P_\theta (t+\Delta t \mid t) \:  A^T + \sigma^2_z]^{-1},\nonumber\\
\hat{\theta}(t+\Delta t \mid & t+\Delta t) = \hat{\theta}(t+\Delta t \mid t)\nonumber \\
& + K_\theta(t+\Delta t) \: \big[Y(t+\Delta t) - \hat{X}(t+\Delta t \mid t) \big],\nonumber\\
P_\theta(t+\Delta t \mid & t+\Delta t) = \big[I-K_\theta(t+\Delta t) \: A \big] \: P_\theta(t+\Delta t \mid t).\nonumber
\end{align}
\end{itemize}
Then  $\hat{X}(t + \Delta t \mid t)$, which is the posterior estimate of $X(t)$, is used to estimate $X(t)$ and similarly, $\hat{\theta}(t + \Delta t \mid t)$ for estimating $\mu_S(t)$ and $ \sigma^2_S(t)$ in~\eqref{eqn_speed_lognormal}.  

\end{document}